\documentclass[twoside]{article}
\usepackage{fleqn,espcrc2}
\usepackage{latexsym}

\newcommand{\AmS}{{\protect\the\textfont2
  A\kern-.1667em\lower.5ex\hbox{M}\kern-.125emS}}
\newcommand{\nn}{\nonumber \\}
\newcommand{\p}[1]{(\ref{#1})}
\newcommand{\be}{\begin{equation}}
\newcommand{\ee}{\end{equation}}
\newcommand{\bea}{\begin{eqnarray}}
\newcommand{\eea}{\end{eqnarray}}

\newcommand{\cA}{{\cal A}}
\newcommand{\cAb}{{\bar{\cal A}}}
\newcommand{\cW}{{\cal W}}
\newcommand{\cWb}{{\bar{\cal W}}}
\newcommand{\lb}[1]{\label{#1}}

\newcommand\q{\quad}
\newcommand\qq{\qquad}
\newcommand\ov{\overline}
\newcommand{\vp}{\varphi}
\newcommand{\bvp}{\bar\varphi}
\newcommand{\bnu}{\bar\nu}

\newcommand{\da}{{\dot{\alpha}}}
\newcommand{\db}{{\dot{\beta}}}

\newcommand{\tta}{\theta_2^\alpha}

\newcommand{\tha}{\theta_3^\alpha}

\newcommand{\btob}{\bar{\theta}^{1\db}}

\newcommand{\bttb}{\bar{\theta}^{2\db}}

\newcommand\ab{{\alpha\beta}}

\newcommand\adb{{\alpha\db}}

\newcommand{\Dot}{D^1_2}
\newcommand{\Dto}{D^2_1}
\newcommand{\Doh}{D^1_3}

\newcommand{\Dht}{D^3_2}
\newcommand{\Dth}{D^2_3}

\newcommand{\Vot}{V^1_2}
\newcommand{\Vto}{V^2_1}
\newcommand{\Voh}{V^1_3}

\newcommand{\Vht}{V^3_2}
\newcommand{\Vth}{V^2_3}

\newcommand{\Doa}{D^1_\alpha}

\newcommand{\Dta}{D^2_\alpha}

\newcommand{\bDta}{\ov{D}_{2\da}}

\newcommand{\bDha}{\ov{D}_{3\da}}

\title{TOWARDS HIGHER-N SUPEREXTENSIONS OF BORN-INFELD THEORY}

\author{E. Ivanov\\\vspace{0.2cm}Bogoliubov Laboratory of Theoretical Physics,
      JINR,\\ 141980 Dubna, Moscow Region, Russia}

\begin{document}

\begin{abstract}
We give a brief account of supersymmetric Born-Infeld theories with
extended supersymmetry, including those with partially broken supersymmetry.
Some latest developments in this area are presented. One of them is $N=3$
supersymmetric Born-Infeld theory which admits a natural off-shell
formulation in $N=3$ harmonic superspace.
\vspace{1pc}
\end{abstract}

\maketitle

\section{INTRODUCTION}

\noindent The main source of current interest in superextensions of
the Born-Infeld (or Dirac-Born-Infeld) action is the fact that these super
BI actions constitute an essential part of the worldvolume actions
of D$p$-branes (see \cite{Ts} and
refs. therein). It is tempting to formulate super BI actions in a manifestly
supersymmetric way using an off-shell superfield approach. In
this approach the worldvolume SUSY is linearly realized, as opposed to the
approach proceeding from a gauge-fixed form of the Green-Schwarz D$p$-actions
\cite{AS}, in which all SUSYs are nonlinear and non-manifest. As was shown
in \cite{BG}, the $N=1, d=4$ super BI theory first formulated in \cite{CF}
in terms of superfields can be re-interpreted as a theory of
partial spontaneous breaking of global $N=2$ SUSY (PBGS) down to $N=1$ SUSY,
with vector $N=1$ multiplet as the relevant Goldstone multiplet. Thus
the PBGS approach provides a powerful way of deriving super BI actions with
extra hidden nonlinearly realized supersymmetries.

This contribution is a brief survey of super BI theories of this kind known
to date. A short description of the recently constructed $N=3$
supersymmetric BI theory \cite{IZ} is also given.

\section{SUPER BI THEORIES AS PBGS SYSTEMS}

\noindent{\it 2.1 $N=2 \rightarrow N=1, d=3$ BI theory}. The basic object of
this simplest super BI theory \cite{ik} is the $N=1, d=3$ spinor gauge
superfield strength $\mu^a(x^{bc}, \theta^d)$, $a,b,c,d =1,2$, subjected
to the constraint
\bea\label{cc13d}
&& D_a\mu^a=0~,
\eea
where $D_a = \partial/\partial \theta^a +1/2\, \theta^b\partial_{ab}$. A
nonlinear extension of the free $N=1, d=3$ Maxwell action, with
the $d=3$ BI action as the bosonic core, reads
\bea
S &\sim & \int d^3x d^2\theta\, w~, \nn
w &\equiv & {1\over 2}\,
\frac{\mu^2}{1 + \sqrt{1 - D^2\mu^2}}~.
\label{muaction}
\eea
Besides being manifestly $N=1$ supersymmetric, this action possesses
the hidden nonlinear SUSY
\bea
&& \delta_\eta \mu_a = \eta_a\left(1- D^2 w\right) + \eta^b\partial_{ab}w~.
\label{Wtrans21}
\eea
The superfield $\mu^a$ is the Goldstone superfield associated with the
spontaneous breaking of the $N=2, d=3$ SUSY,
\be
\{Q_a, Q_b \} = \{S_a, S_b \} = P_{ab}~, \quad \{Q_a, S_b \} = 0~,
\ee
down to the $N=1$ one generated by $Q_a, P_{ab}$. The Goldstone fermion
$\psi^a$ associated with the spontaneously broken generators $S_a$ in the
standard nonlinear realizations approach (with making use of the
exponential parametrization of the Poincar\'e supergroup elements) is
related to $\mu^a$ by
\bea
&& \psi^a = \frac{\mu^a}{1- D^2 w}~. \lb{psimu}
\eea
In components, \p{muaction} gives rise to a static-gauge form of the
space-filling D2-brane action.

Now let us briefly outline how the super BI action \p{muaction} can be deduced.
The derivation follows a generic method \cite{ika,dik} which is also
applicable to other PBGS cases \cite{lech,ei}.

The starting point is the appropriate linear realization of the considered
PBGS pattern. It is obtained by embedding the $N=1, d=3$ Maxwell superfield
strength $\mu_a$ into a linear $N=2, d=3$ multiplet. The latter should have
such a transformation law under the $S$-supersymmetry that  $\mu_a$
transforms with an inhomogeneous term $\sim \eta_a$ and so can be
interpreted as the Goldstone fermion of  linear realization.

The appropriate $N=2, d=3$ supermultiplet was proposed in \cite{BZ} as a
deformation of the $N=2, d=3$ Maxwell multiplet. The superfield constraints
defining this deformed multiplet  can be taken in the form \cite{ei}
\bea
&& \mbox{(a)}\;\;\left[ (D)^2 - (D^\zeta)^2 \right] W = -2i~, \nn
&& \mbox{(b)} \;\; D^{a}D^{\zeta}_a W = 0~, \label{constrW}
\eea
where $W(x,\theta, \zeta)$ is a real $N=2, d=3$ superfield ($\zeta_a$ is
an extra Grassmann coordinate). The standard $S$-supersymmetry
transformation law of $W$,
\bea
&& \delta_\eta W = -\eta^a\left( \frac{\partial}{\partial \zeta^a} -{1\over
2}\zeta^b\partial_{ab} \right) W~,  \label{Wtrnas}
\eea
implies the following transformation laws for the irreducible
$N=1$ superfield components of $W(x,\theta, \zeta)$, $\mu_a \equiv
-iD^\zeta_a W|_{\zeta=0}$ and $w \equiv i/2\, W|_{\zeta=0}$,
\bea
&& \mbox{(a)} \;\;\delta_\eta \mu_a = \eta_a\left(1- D^2 w\right) +
\eta^b\partial_{ab}w~, \nn
&& \mbox{(b)}\;\;\delta_\eta w = {1\over 2}\,
\eta^a\mu_a~. \label{Wtrans2}
\eea
It is easy to check that eq. (\ref{Wtrans2}a) is consistent with the
Bianchi identity \p{cc13d} (which is none other than eq. (\ref{constrW}b)).

The additional homogeneously transforming $N=1$ superfield $w(x,\theta)$
can be expressed in terms of the Goldstone-Maxwell
one $\mu_a$ by enforcing nonlinear constraints the precise form of which
is dictated by the generic method of refs. \cite{ika}-\cite{ei} applied
to the given system.

As the first step, one defines superfields $\tilde{\mu}_a$ and
$\tilde{w}$ as finite  $S$-supersymmetry transforms of $\mu_a$ and $w$,
with the transformation parameter $\eta^a$ being replaced by
$-\psi^a(x,\theta)$:
\bea
\tilde{\mu}_a &=& \mu_a - \psi_a\left(1- D^2 w\right) -
\psi^b\partial_{ab}w \nn
&& \,-{1\over 4} \psi^2 \partial_{ab}\mu^b ~, \nn
\tilde{w} &=& w - {1\over 2}\,\psi^a\mu_a + {1\over 4}\,\psi^2 \left(1-
D^2w\right)~. \lb{d3rel}
\eea
These quantities homogeneously transform with respect to the
whole $N=2, d=3$ SUSY. Therefore, one can covariantly equate them to
zero,
\bea
&& \tilde{\mu}_a = \tilde{w} = 0~. \lb{d3eqs}
\eea
From these covariant constraints one derives the equivalence relation
between $\psi^a$  and $\mu^a$ \p{psimu}, as well as  the relation
\bea
&& w = {1\over 4}\,\frac{\mu^2}{1- D^2 w}~.
\eea
These are precisely the equations postulated in \cite{ik}
(up to a rescaling of $w$). They can be used to  express $w$ in terms
of either $\psi^a$ or $\mu^a$
\bea
w = {1\over 4}\, \frac{\psi^2}{1 + {1\over 4} D^2 \psi^2} =
 {1\over 2}\,\frac{\mu^2}{1 + \sqrt{1 - D^2\mu^2}}~. \label{D2dens}
\eea
This composite superfield is just the Goldstone superfield
Lagrangian density in the action \p{muaction}.

The same superfield D2-brane action can be written in a manifestly
$N=2$ supersymmetric  form as an integral over the whole $N=2$ superspace,
with either $W^2$ or $N=2, d=3$ Fayet-Iliopoulos term as the Lagrangian
densities.
\vspace{0.3cm}

\noindent{\it 2.2 $N=2 \rightarrow N=1, d=4$ BI theory}. This case
\cite{BG} corresponds to the partial breaking of $N=2, d=4$ SUSY,
\bea
&& \left\{ Q_{\alpha},{\bar Q}_{\dot\alpha} \right\}=
\left\{ S_{\alpha}, {\bar S}_{\dot\alpha}
\right\}=2P_{\alpha\dot\alpha}~, \nn
&& \left\{ Q_{\alpha}, S_{\beta} \right\} =0~, \label{susyD43d}
\eea
down to the $N=1, d=4 \propto (Q, \bar Q,
P)$, with a vector $N=1, d=4$ multiplet as the Goldstone one. In terms of
the $N=1, d=4$ Maxwell superfield strength $W^\alpha (z), \bar
W^{\dot\alpha}(z)$ ($z = (x^{\alpha\dot\alpha}, \theta^\alpha, \bar
\theta^{\dot\alpha})$),
\bea
&& \bar D^{\dot\alpha} W^\beta =0~, \; D^\alpha W_\alpha +\bar
D_{\dot\alpha}\bar W^{\dot\alpha} =0~, \label{N1Max}
\eea
the Goldstone superfield action ($N=1, d=4$ BI action) reads
\bea
&& S \sim \int d^4 x_L d^2\theta\, \phi~, \nn
&& \phi  = W^2 + {1\over 2}\bar D^2
\frac{W^2\bar W^2}{1
-{1\over 2}A + \sqrt{1 - A +{1\over 4} B^2}}~, \nn
&& A \equiv {1\over 2}\left(D^2W^2 +\bar D^2\bar W^2\right)~, \nn
&& B \equiv
{1\over 2}\left(D^2W^2 -\bar D^2\bar W^2\right)~. \label{N2BI}
\eea
The equivalence relation between the canonical nonlinear realization
Goldstone spinor superfields $\psi^\alpha, \bar \psi^{\dot\alpha}$
and $W^\alpha, \bar W^{\dot\alpha}$ is \cite{ei}
\bea
&& W^\alpha = \psi^\alpha \left(1 -{1\over 4}\bar D^2 \bar\phi\right)
+ \ldots~,
\eea
where $\ldots$ stand for terms with $x$-derivatives. The nonlinear SUSY
acts as
\bea
&& \delta_\eta W_\alpha = \eta_\alpha \left(1-{1\over 4} \bar D^2 \bar \phi
\right) - i \bar\eta^{\dot\alpha} \partial_{\alpha\dot\alpha}\phi~.
\eea
In components, \p{N2BI} describes the space-filling D3-superbrane
in a static gauge.

The action \p{N2BI} can be deduced \cite{ei} from the appropriate
linear realization of the PBGS pattern  $N=2 \rightarrow N=1, d=4$,
following the procedure similar to that applied in the
$N=2 \rightarrow N=1, d=3$ case above. The starting point is $N=2, d=4$
Goldstone-Maxwell multiplet in the
$N=2$ superspace $z = (x^{\alpha\dot\alpha},\theta^\alpha_i,
\bar\theta^{\dot\alpha i})$. It is defined by the following
deformation \cite{IZ1} of the standard $N=2$ Maxwell superfield strength
constraints
\bea
&& \mbox{(a)} \;\;D^{ik}{\cal W} -\bar{D}^{ik}\bar {\cal W} = i M^{(ik)}~,
\nn
&& \mbox{(b)} \;\;D^i_\alpha \bar{\cal  W} = \bar D_{\dot\alpha i} {\cal W}
=0~.
\label{deformN2}
\eea
Here,
$$
D^i_\alpha  = \frac{\partial}{\partial \theta^\alpha_i} -i
\bar\theta^{\dot\alpha i}\partial_{\alpha\dot\alpha}~, \;
\bar D_{\dot\alpha i} =
-\frac{\partial}{\partial \bar\theta^{\dot\alpha i}}
+i \theta^{\alpha}_i \partial_{\alpha\dot\alpha}~,
$$
$$
D^{ik} = D^{\alpha i}D^k_\alpha~, \; \bar D^{ik} =
\bar D_{\dot\alpha}^i \bar
D^{\dot\alpha k}~,
$$
and $M^{ik} = M^{ki}$ is a triplet of constants which
explicitly break the automorphism  $SU(2)_A$ of $N=2$ supersymmetry down to
$U(1)_A$ and satisfy the pseudo-reality  condition
$$
\overline{(M^{ik})} = \epsilon_{in}\epsilon_{km}M^{nm}~.
$$

Now we pass to the $N=1$ superfield notation by relabelling the Grassmann
coordinates and spinor derivatives as
$$
\theta^\alpha_1 \equiv \theta^\alpha~, \;\theta^\alpha_2 \equiv \zeta^\alpha,
\; D^1_\alpha \equiv D_\alpha, \; D^2_\alpha \equiv D^\zeta_\alpha~.
$$
In order to have the off-shell $S$-supersymmetry (acting as
$\zeta$-supertranslations) spontaneously broken while the $Q$-supersymmetry
unbroken, we choose the following frame with respect to the
explicitly broken $SU(2)_A$ \bea
&& M^{12} = 0~, \quad M^{11} = M^{22} = m~,
\eea
where $m$ is a real constant. Like in the case of D2-brane it is fixed up to
rescaling of  $W$. A convenient choice is
$$
m = -2~.
$$
It will be also convenient to choose a basis in $N=2$ superspace where the
chirality with respect to the variable $\zeta^\alpha$ is manifest
\bea
&& \bar D^\zeta_{\dot\alpha} = -\frac{\partial}{\partial
\bar\zeta^{\dot\alpha}}~, \quad   D^\zeta_{\alpha} =
\frac{\partial}{\partial
\zeta^{\alpha}} -2i \bar\zeta^{\dot\alpha}\partial_{\alpha\dot\alpha}~.
\eea
In this basis, constraints \p{deformN2} imply the following structure of
the superfield ${\cal W}(x, \theta, \zeta)$
\bea
&& {\cal W} = -{i\over 2}\phi + i\zeta^\alpha W_\alpha
-i {1\over 2} \zeta^2\left(1 - {1\over 4}
\bar D^2 \bar\phi\right), \nonumber
\eea
where $\phi$ and $W_\alpha$ are chiral $N=1$ superfields
\bea
&& \bar D_{\dot\alpha} \phi = \bar D_{\dot\alpha} W_\alpha = 0~,
\eea
and the fermionic superfield $W_\alpha$ obeys the $N=1$
Maxwell superfield strength constraint \p{N1Max}.

The $S$-supersymmetry transformation of the $N=2$ superfield ${\cal W}$
\bea
\delta_\eta {\cal W} = -\left[\eta^\alpha \frac{\partial}{\partial
\zeta^\alpha} + \bar\eta^{\dot\alpha}\left(\frac{\partial}{\partial
\bar\zeta^{\dot\alpha}} +2i \zeta^\alpha
\partial_{\alpha\dot\alpha}\right)\right]{\cal W} \nonumber
\eea
implies the following ones for its $N=1$ superfield components $\phi$ and
$W_\alpha$
\bea
&& \delta_\eta \phi = 2(\eta W)~, \quad \delta_\eta \bar\phi = 2(\bar
W\bar\eta)~, \nn
&&\delta_\eta W_\alpha = \eta_\alpha \left(1-{1\over 4} \bar D^2 \bar \phi
\right) - i \bar\eta^{\dot\alpha} \partial_{\alpha\dot\alpha}\phi~, \nn
&& \delta_\eta \bar W_{\dot\alpha} = \overline{(\delta_\eta W_\alpha)}~.
\label{tranfiW}
\eea
The superfield $W_\alpha$ shows up an inhomogeneous shift $\sim \eta_\alpha$
in its transformation, so it is the Goldstone fermion of the  linear
realization of the considered $N=2 \rightarrow N=1, \;d=4$ PBGS pattern
(Goldstone-Maxwell $N=1$ superfield).

After this one can start the algorithmic procedure of
passing to the relevant nonlinear realization.
It goes in full analogy with the $N=2 \rightarrow N=1, d=3$ case. Firstly
one constructs finite $\eta$-transformations of the superfields
$\phi$  and $W_\alpha$ proceeding from the infinitesimal ones \p{tranfiW}.
As the next step, one replaces, in the transformed superfields, the
transformation parameters by the original nonlinear realization Goldstone
fermions, $\eta_\alpha \rightarrow -\psi_\alpha, \;\bar\eta_{\dot\alpha}
\rightarrow -\bar\psi_{\dot\alpha}$. At last, one imposes the covariant
constraints on the so defined superfields
\bea
&& \tilde\phi  = \tilde W_\alpha = 0~, \label{basicD3}
\eea
and obtains the appropriate relations between $\phi, W_\alpha$ and
$\psi_\alpha$. After some algebra, one ends up with the simple relations
\be
\phi = \frac{W^2}{1 - {1\over 4} \bar D^2 \bar\phi}~, \quad \bar\phi =
\frac{\bar W^2}{1 - {1\over 4} D^2 \phi}~, \label{basicRel}
\ee
which, up to a rescaling of $\phi$, are just those postulated in \cite{BG}
and derived from the nilpotency condition in \cite{RT}. Their solution is
the expression for $\phi$ in \p{N2BI}.

Note that the  Goldstone superfield Lagrangian density for the
$N=2 \rightarrow N=1$ PBGS in \p{N2BI} can be equivalently rewritten
in $N=2, d=4$ superspace in terms of the constrained $N=2$ gauge superfield
strength ${\cal W}, \bar{\cal W}$ as the Fayet-Iliopoulos term or as the
kinetic term of $N=2$ Goldstone-Maxwell multiplet.
\vspace{0.3cm}

\noindent{\it 2.3 $N=4 \rightarrow N=2, d=4$ BI theory}. The $N=2, d=4$ super
BI action can be constructed in terms of the $N=2$ Maxwell superfield
strength ${\cal W}(z),
\bar{\cal W}(z)$, $z = (x^{\alpha\dot\alpha},\theta^\alpha_i,
\bar\theta^{\dot\alpha i})$, defined by the constraints
\bea
&& \bar D^i_{\dot\alpha} {\cal W} = D_i^{\alpha} \bar{\cal W} =0~, \quad
D^{ik}{\cal W} = \bar D^{ik}\bar{\cal  W}~, \label{oshN2M}
\eea ($i,k = 1,2$), which is the $M^{ik} = 0$ version of the constraints
\p{deformN2}. The simplest $N=2$ BI action was proposed in \cite{Ke}.
It goes into the previously given $N=1$ BI action after the appropriate
reduction, but it does not possess a second nonlinearly realized $N=2$ SUSY.
So it cannot be interpreted as the Goldstone superfield action for the
PBGS option $N=4 \rightarrow N=2$ in $d=4$.

A recursive procedure of restoring the correct action was
developed in \cite{bik1,bik2}. The point of departure is the
following $N=4, d=4$ superalgebra
\bea
&& \left\{
Q_{\alpha}^i, {\bar Q}_{\dot{\alpha}j}
        \right\}=\left\{ S_{\alpha}^i, {\bar S}_{\dot{\alpha}j} \right\}
=2\delta^i_jP_{\alpha\dot{\alpha}}\;, \nn
&& \left\{ Q_{\alpha}^i,  S_{\beta}^j
   \right\}=2\varepsilon^{ij}\varepsilon_{\alpha\beta}Z~. \label{n4sa}
\eea The basic Goldstone superfield supporting the $1/2$ breaking of this
$N=4$ SUSY is a scalar one $W, \bar W$ associated with the complex central
charge generator $Z, \bar Z$. It obeys a nonlinear version of the
constraints \p{oshN2M} and is related to ${\cal W}, \bar{\cal W}$ by a
complicated equivalence redefinition. Up to the sixth order, the $N=4
\rightarrow N=2$ Goldstone-Maxwell superfield action and transformations of
the central charge and second hidden SUSY (with the generators $S, \bar S$)
read
\bea
\delta {\cal W} &=& f -{1\over 2}\bar D^4 (f A) + {1\over 4} \Box
(\bar f \bar A) \nn && +\, {1\over 4i}\,\bar D^{i\dot\alpha}\bar f
D^{\alpha}_i\,\partial_{\alpha\dot\alpha}\bar A~, \label{4transf} \\
A &=& \bar{\cal W}^2\left(1 + {1\over 2}D^4{\cal W}^2 \right), \;
D^4 = \frac{1}{48}
D^{ik}D_{ik}~, \nn
f &=& c+ 2i\,\eta^{i\alpha}\theta_{i\alpha}\,, \label{fdef} \\
S^{(6)}_{bi} &\sim & \int d\zeta_L {\cal W}^2
+ \mbox{c.c.}+ {1\over 2}\int d z \{\; {\cal W}^2\bar{\cal W}^2  \nn
&& \times\,\left[2 + \left( D^4{\cal W}^2 + \bar D^4 \bar{\cal
W}{}^2\right)\right]   \nn
&& -\, {1\over 9} {\cal W}^3 \Box \bar{\cal
W}^3 + O(W^8)\;\}~, \label{action}
\eea
where $dz$ and $d\zeta_L$ are measures of integration
over the whole $N=2$ superspace and its chiral subspace.
In \cite{bik2} the action was restored up to 10th
order using some algorithmic iteration procedure. The full action contains
the standard BI action and some disguised form of the Nambu-Goto action for
two physical scalar fields. It is a gauge-fixed action of D3-superbrane in
$D=6$.

Let us say a few words on the aforementioned recursive procedure of restoring
the full superfield action and its relation to the approach based on the
linear realizations of PBGS exemplified in the previous two subsections.

As was proposed in \cite{bik2}, the linear Goldstone-Maxwell multiplet
relevant to the case at hand is an infinite-dimensional linear multiplet of
$N=4$ superalgebra \p{n4sa}, with a non-trivial realization of the central
charges $Z$, $\bar Z$. To be more precise, one embeds the $N=2$ Maxwell
superfield strength  ${\cal W}$ into an infinite-dimensional  $N=4$ multiplet
\bea
&& {\cal W},\;\; \bar{\cal  W},\;\; {\cal
A}_n, \;\; \bar{\cal A}_n, \qquad (n = 0, 1, 2, \dots)~,
\eea
where ${\cal A}_n$ are chiral (otherwise unconstrained) $N=2$
superfields,
\bea
&& \bar D_{\dot\alpha i} {\cA_n} =0~,\;
D_{\alpha}^i \bar{\cA_n} =0~. \label{ivan-defA}
\eea
The following transformations
\bea
\delta {\cal W} &=& f -{1\over 2}\bar D^4 (f \cAb_0 ) +
{1\over 4} \Box (\bar f \cA_0) \nn
&& +\, {1\over 4i}
\,\bar D^{i\dot\alpha}\bar f D^{\alpha}_i\,\partial_{\alpha\dot\alpha}\cA_0~,
\label{ivan-4transf4d} \\
\delta {\cA_0} &=& 2f\cW  +
{1\over 4}\bar f \Box \cA_1 + {1\over 4i}
\,\bar D^{i\dot\alpha}\bar f D^{\alpha}_i\,
\partial_{\alpha\dot\alpha} \cA_1, \nn
\delta {\cA_1} &=& 2f \cA_{0}  +
{1\over 4}\bar f \Box \cA_{2} + {1\over 4i}
\,\bar D^{i\dot\alpha}\bar f D^{\alpha}_i\,\partial_{\alpha\dot\alpha}
\cA_{2} \nn
&& ........... \nn
\delta {\cA_{n}} &=& 2f \cA_{n-1}  +
{1\over 4} \bar f \Box \cA_{n+1} \nn
&& +\, {1\over 4i}
\,\bar D^{i\dot\alpha}\bar f D^{\alpha}_i\,
\partial_{\alpha\dot\alpha} \cA_{n},\quad (n\geq 1)\label{ivan-N4b}
\eea
where the function $f$ was defined in \p{fdef}, close off shell
both among  themselves and with those of the manifest $N=2$
supersymmetry  just according to the superalgebra
\p{n4sa}. The central charge ($c, \bar c$) transformations non-trivially
act on this infinite tower of $N=2$ superfields.

A good candidate for the chiral $N=2$ Lagrangian density
is the superfield $\cA_0$. Indeed, the ``action''
\bea
&& S=\int d^4x d^4\theta \cA_0 + \int d^4x d^4\bar\theta \cAb_0
\label{ivan-action1}
\eea
is invariant with respect to the transformation \p{ivan-N4b}
up to surface terms.

It remains to define covariant constraints
which would express $\cA_0$,  $\cAb_0$ in terms
of ${\cal W}$, $\bar{\cal W}$, with preserving the linear representation
structure \p{ivan-4transf4d}, \p{ivan-N4b}. Because of an infinite number
of $N=2$ superfields ${\cal A}_n$, there should exist an infinite set of
constraints expressing  these superfields  through the basic Goldstone
ones ${\cal W}$, $\bar{\cal W}$. The procedure of deducing this set of
constraints was described in \cite{bik2}. The first two constraints read
\bea\label{ivan-constr1}
\phi_0 &=& \cA_0 \left( 1-\frac{1}{2}{\bar D}{}^4\cAb_0\right) -
\cW^2 \nn
&& -\, \sum_{k=1}\frac{ (-1)^k}{2\cdot 8^k}
   \cA_k\Box^k {\bar D}{}^4 \cAb_k = 0~,  \nn
\phi_1 &=& \Box \cA_1 +2\left( \cA_0\Box \cW - \cW\Box\cA_0 \right) \nn
&& - \sum_{k=0}\frac{ (-1)^k}{2\cdot8^k}  \left( \Box \cA_{k+1}\Box^k{\bar
D}{}^4 \cAb_k \right. \nn
&& \left. -\, \cA_{k+1} \Box^{k+1} {\bar D}{}^4 \cAb_k \right) =0\;,
\eea
and so on.

At present we do not know how to explicitly solve
this set of constraints and to find a closed
expression for the Lagrangian densities ${\cal A}_0$, $\bar{\cal A}_0$.
We can only recover the general solution by iterations.
E.g.,
\bea
&& \cA_0= \cW^2 + \cA_0^{(4)}+\cA_0^{(6)}+\cA_0^{(8)}+\ldots \; ,\nn
&&\cA_0^{(4)}= \frac{1}{2}\cW^2 {\bar D}{}^4\cWb^2~, \; \ldots ~.
\eea
The action, up to the 10th order in  ${\cal W}, \bar{\cal W}$, was found in
\cite{bik2}. It turned out to coincide, at least up to the 8th order,
with the action deduced in \cite{KT} from the
requirements of  self-duality and invariance under nonlinear shifts of ${\cal
W}, \bar{\cal  W}$ (the $c, \bar c$ transformations in our notation).  This is
an indication that the full $N=4 \rightarrow N=2$ BI action is also
self-dual like its $N=2 \rightarrow N=1$ prototype \cite{BG}.
\vspace{0.3cm}

\noindent{\it 2.4 $N=8 \rightarrow N=4, d=4$ BI theory}. No manifestly
$N=4$ supersymmetric off-shell actions are known for $N=4, d=4$ Maxwell theory,
so no such actions can be defined for the BI deformations of the latter. The
best what one can hope to gain is $N=4$ BI actions with the $N=1$,
$N=2$ \cite{Ts} or at most $N=3$ \cite{IZ} manifest off-shell SUSYs.
However, it is still possible to derive the superfield equations of motion
of $N=4$ BI theory, in a manifestly $N=4$ supersymmetric form and with one extra
nonlinearly realized $N=4$ SUSY, within the nonlinear realizations formalism
applied to the following $N=8, D=4$ superalgebra \cite{bik1}:
\bea
&& \left\{ Q_{\alpha}^i, {\bar Q}_{\dot{\alpha}j}
        \right\}=\left\{ S_{\alpha}^i, {\bar S}_{\dot{\alpha}j}\right\}
=2\delta^i_jP_{\alpha\dot{\alpha}}\;, \nn && \left\{ Q_{\alpha}^i,
S_{\beta}^j\right\}=\varepsilon_{\alpha\beta} Z^{ij} \label{n8sa} \\ &&
{\bar Z}_{ij}=\left(Z^{ij}\right)^* =
\frac{1}{2}\varepsilon_{ijkl}Z^{kl}\;.\label{reality} \eea The basic
Goldstone superfield supporting partial breakdown of this SUSY down to $N=4,
d=4$ SUSY $ \propto (Q, P)$ is an $N=4$ superfield $W_{ik}~, \bar W^{ij} =
\frac{1}{2}\varepsilon^{ijkl} W_{kl}~,$ associated with the generator
$Z^{ik}$. One also introduces spinor Goldstone superfields
$\psi^{\alpha}_i, {\bar\psi}^{\dot\alpha i}$ associated with $S_{\alpha}^i$
and ${\bar S}_{\dot{\alpha}j}$. The covariant superfield equations of the
$N=8 \rightarrow N=4$ super BI theory read \bea \mbox{(a)} \quad \psi_{\alpha
i}+ \frac{2i}{3}{\cal D}_{\alpha}^jW_{ij} = 0, &&\nn
\mbox{(b)}\quad {\cal
D}_{\alpha}^k W_{ij} - \frac{1}{3} \left[\delta_i^k {\cal D}_{\alpha}^m
W_{mj} - (i\leftrightarrow j)\right] = 0 &&
\label{eqdyn}
\eea
(plus their c.c.). Here ${\cal D}_{\alpha}^j, \bar{\cal D}_j^{\dot\alpha}$
are the appropriate covariantization of the flat $N=4$ spinor derivatives.
They nonlinearly depend on $\psi^{\alpha}_i, {\bar\psi}_i^{\dot\alpha}$ which
can be covariantly expressed through derivatives of $W_{kl}$ by eq.
(\ref{eqdyn}a). Eq. (\ref{eqdyn}b) is a covariantization of the standard
superfield constraints of the on-shell $N=4$ Maxwell theory \cite{N4con},
and it contains the dynamical equations for the component fields. One of them
is a disguised form of the BI equations. The full set of equations is
a manifestly worldvolume supersymmetric form of the equations of the gauge-fixed
D3-superbrane in $D=10$,  with 6 physical bosonic fields of $N=4$
Goldstone-Maxwell multiplet being transverse brane coordinates.

As was already mentioned, one cannot hope to construct an off-shell
$N=4$ supersymmetric superfield action for this super BI system, since
no such action exists even for the ordinary $N=4$ gauge theory. However, for
the first non-trivial term in the BI action, the quartic term $\sim
F^{\alpha\beta}F_{\alpha\beta}\bar F^{\dot\alpha\dot\beta}
\bar F_{\dot\alpha\dot\beta}$,
$N=4$ completions with $N=1$ and $N=2$ off-shell supersymmetries are
known. These were given, respectively, in terms of $N=1$ superfields
\cite{Ts} and $N=2$ projective superfields \cite{GRo}. Here we present
this completion in terms of off-shell $N=2$ harmonic
superfields \cite{ivunp}.

In the harmonic superspace (HSS) description \cite{GIK1,book}, $N=4$
vector multiplet is represented by the standard gauge $N=2$ superfield
strength ${\cal W}(z), \bar{\cal W}(z)$ and the analytic
hypermultiplet superfield $q^{+a}(\zeta, u)$. Here $\{\zeta, u\}
\equiv \{x^{\alpha\dot\beta}, \theta^{+\alpha}, \bar\theta^{+\dot\alpha},
u^{\pm i} \}$ are co-ordinates of an analytic subspace of $N=2$ harmonic
superspace $\{z, u^{\pm i} \}$ and $u^{\pm i}, u^{+ i}u_i^- = 1$, are harmonic
variables parametrizing some internal 2-sphere $S^2$. The indices $a$ and $i$
are doublet indices of two mutually commuting $SU(2)$ groups, $a =1,2; i = 1,2$.
Further details can be found in \cite{GIK1,book}. The free $N=4$
Maxwell theory action is given by
\bea
S^{N=4}_{free} &=& {1\over 8}\,\int d\zeta_L {\cal W}^2
+ \mbox{c.c.} \nn
&& -\,
\frac{1}{2}\int d\zeta^{(-4)} q^{+a}
D^{++}q^+_a~,
\label{4}
\eea
where $D^{++}$ is the analyticity-preserving harmonic derivative
and $d\zeta^{(-4)}$ is the measure of integration over the analytic superspace.
Besides being manifestly $N=2$ supersymmetric, the action \p{4} is invariant
under one more $N=2$ SUSY which forms $N=4, d=4$ SUSY together
with the manifest $N=2$ one. For our purposes it suffices to know only the
on-shell form of transformations of this hidden SUSY:
\bea
&& \delta{\cal  W} = {1\over 2}\bar\epsilon^{\dot\alpha a}\,\bar
D^-_{\dot\alpha}q^+_a\;, \quad
\delta \bar{\cal W} = {1\over 2}\epsilon^{\alpha a}\,
D^-_{\alpha}q^+_a~, \nn
&& \delta q^\pm_a ={1\over 4}\,(\epsilon^\beta_a D^\pm_\beta{\cal W} +
\bar\epsilon^{\dot\alpha}_a\bar D^\pm_{\dot\alpha} \bar{\cal W})~,
\label{onshell}
\eea
where $\epsilon^{\alpha a}, \bar \epsilon^{\dot\alpha a}$ are the Grassmann
transformation parameters, $D^{\pm}_\alpha, \bar D^\pm_{\dot\alpha}$ are
harmonic projections of the flat $N=2$ spinor derivatives and $q^{-a} \equiv
D^{--}q^{+a}$, $D^{--}$ being the second harmonic derivative. The quartic
superfield term which is invariant under the transformations \p{onshell}
and yields the correct quartic term $\sim F^2\bar F^2$ in the component
BI action (with the correct relative coefficient w.r.t. the free Maxwell
action) is uniquely restored, up to terms vanishing on the free mass shell,
\bea
&& S_4^{N=4} = {1\over 32}\int dzdu\, \{\, {\cal W}^2\bar{\cal W}^2 - 4(q^+\cdot
q^-)\nn
&& \times\, [{\cal W}\bar{\cal W}  - {1\over 3}(q^+\cdot q^-)]\,\},
\label{N44}\\
&& (q^+\cdot q^-)
\equiv q^{+a}q^-_a~. \nonumber
\eea
Here $du$ is the measure of integration over harmonics ($\int du\cdot 1 = 1$).
The problem of $N=4$-completing of higher-order terms of the $N=2$ BI action
(even of its simplest version \cite{Ke}) is technically very
complicated because the form of the hidden on-shell $N=4$ transformations
\p{onshell} is modified from order to order in superfields.

It is remarkable that in $N=3$ HSS \cite{GIK2} one can construct
an off-shell $N=3$ superextension of the {\it full} BI action \cite{IZ}.

\section{N=3 BORN-INFELD THEORY}
\noindent{\it 3.1 Introduction}. Since the $N > 2$ super BI actions
are extensions of the corresponding super Maxwell actions,
the necessary condition of the existence of some off-shell super BI action
is the existence of such an
action for the relevant free super Maxwell theory.
The maximally supersymmetric off-shell
formulation of $N=4$ gauge theory is that with manifest $N=3$ supersymmetry.
It was given in \cite{GIK2} in the framework of $N=3$ harmonic superspace (HSS).

In \cite{IZ}, starting from this formulation, we have constructed an
$N=3$ superextension of the full BI action. As distinct from the previously
known $N=1$ and $N=2$ super BI actions, the construction of the $N=3$ BI action
is by no means a straightforward order-by-order supersymmetrization of the
bosonic BI action. The main novel feature stems from the crucial property that
the Grassmann-analytic gauge potentials of $N=3$
gauge theory in $N=3$ HSS \cite{GIK2} contain, besides the physical fields
including  the standard gauge potential $A_m$, also an infinite number
of the auxiliary fields. Among them there is an
independent bispinor field $H_{\alpha\beta} =H_{\beta\alpha} $. The correct
bilinear Maxwell term in the component action arises only after
elimination of this field by its algebraic equation of motion.
The $N=3$ gauge superfield strength  contains the combination
$V_{\alpha\beta}={1\over4}[H_{\alpha\beta}+ F_{\alpha\beta}(A)]$ of the
auxiliary field and the gauge field strength.

The auxiliary component $V_{\alpha\beta}$ can be interpreted as
a Legendre-type transform variable for the gauge field strength
$F_{\alpha\beta}(A)$. It turns out that this specific Legendre transform
of the standard bosonic BI action is
determined by a real function $E$ of the single variable $a= V^2\bar{V}^2$
where $V^2=V^{\alpha\beta} V_{\alpha\beta}$. The problem of $N=3$
supersymmetrization of the BI action is then reduced to the construction
of self-interaction superfield terms of the order 4k in the auxiliary field
$V_{\alpha\beta}$. All these terms can be
constructed as the appropriate powers of the off-shell $N=3$ superfield
strengths and their spinor derivatives in the framework of an analytic
subspace of $N=3$ HSS.
A generic function $E(a)$ exhausts the complete set of the $SO(2)$ self-dual
nonlinear extensions of the Maxwell action, the BI one being a special
representative of them. All such actions can be $N=3$
supersymmetrized off shell.
\vspace{0.3cm}

\noindent{\it 3.2 Elements of $N=3$ harmonic formalism}. The fundamental objects
of the abelian $N=3$ gauge theory are three harmonic gauge potentials living
as unconstrained superfields on the $(4{+}6|8)$-dimensional analytic
subspace $H(4{+}6|8) =\{\zeta, u\}$ of $N=3$ HSS
\bea &&\Vot(\zeta,u)~,\q\Voh(\zeta,u)~,\q \Vth~,\nn
&&\Vot=-\widetilde{(\Vth)}~,\qq\Voh=\widetilde{(\Voh)}~.\lb{C2}
\eea
The definition of the generalized conjugation $\sim $ preserving $N=3$ Grassmann
harmonic analyticity and the precise content of the analytic coordinate  set
$\{\zeta, u \}$ can be found in \cite{GIK2,IZ}.
The potentials undergo abelian  gauge transformations with a real analytic
parameter $\lambda(\zeta,u)$:  \be
\delta\Vot=i\Dot\lambda~,
\q\delta\Voh=i\Doh\lambda~,\q\delta\Vth=i\Dth\lambda~. \lb{C3}
\ee
The potential $V^1_3$ can be consistently expressed in terms of
the two remaining ones by imposing the conventional constraint
\bea
&& \hat{V}^1_3\equiv \Dot\Vth-\Dth\Vot~.
\eea
The free $N=3$ gauge theory action has the following form:
\bea
S_2(\Vot,\Vth) &=& -{1\over4f^2}\int d\zeta(^{33}_{11}) du
[\,\Vth\Doh\Vot \nn
&& +\, {1\over2}(\Dot\Vth-\Dth\Vot)^2\,]~,
\lb{freehs}
\eea
where the analytic superspace integration
measure  $d\zeta(^{33}_{11})du= d^4x_A d^8\theta_A(^{33}_{11})du$ is
defined in \cite{GIK2,IZ} and we have introduced the coupling constant
$f$ of dimension $-2$, so that $[V^1_2]= -2$ and the gauge field
strength is dimensionless. Besides an infinite number of gauge components
accounted for by the
gauge freedom \p{C3}, the gauge potentials possess an infinite number of
the auxiliary field components. The latter disappear only on the mass shell
defined by the free equations of motion following from \p{freehs}:
\bea
&& F^{11}_{23}=\Doh \Vot -\Dot \hat{V}^1_3 =0~, \nn
&& F^{12}_{33}=\Dth \hat{V}^1_3 -\Doh \Vth =0~.\lb{C5}
\eea

For our further purposes it will be important to know the full
structure of the bosonic $SU(3)$ singlet sector in the component
expansion of the off-shell analytic potentials $\Vot$ and $\Vth$
in the WZ gauge. A simple analysis yields
\bea
&& v^1_2=\tta\btob A_\adb +i(\theta_2)^2\bar\theta^{1(\da}
\bar\theta^{2\db)} \bar{H}_{\da\db} \nn
&& +\, i(\theta_2)^2(\bar\theta^1 \bar\theta^2)
C~, \lb{12} \\
&& v^2_3=\tha\bttb A_\adb
-i\theta_2^{(\alpha}\theta_3^{\beta)}(\bar\theta^2)^2H_{\alpha\beta} \nn
&& -\,i(\theta_2\theta_3)(\bar\theta^2)^2C,
\lb{singl}
\eea
where $H_{\alpha\beta} = H_{\beta\alpha}~, \bar C = C$, and the spinor
representation for the gauge field strength was used
\bea
&&F_\ab(A)\equiv \partial_{(\alpha}^\db A_{\beta)\db}~,\;
\bar{F}_{\da\db}(A)\equiv \partial^\beta_{(\da}A_{\beta\db)}~,
\nn
&& \partial_\alpha^\db \bar F_{\db\da} -\partial^\beta_\da F_{\beta\alpha}
= 0~. \lb{Bian}
\eea
We observe that the auxiliary dimensionless symmetric tensor
and scalar fields
$H_\ab$ and $C$ are present in the off-shell $SU(3)$  singlet
sector in parallel with the gauge potential $A_{\alpha\dot\beta}$
and its covariant field strength. The fields
$H_{\alpha\beta}, \bar H_{\da\db}$ play a crucial role
in constructing $N=3$ supersymmetric BI action.

The gauge fields part of the off-shell super $N=3$ Maxwell component
Lagrangian corresponding to \p{freehs} is
\bea
&& L_2(F,H,C)={1\over16f^2}[\,H^2+\bar{H}^2 \nn
&& -\,6\,(\bar{H}\bar{F}+ HF) + F^2+\bar{F}^2 + 8 C^2 \,]~. \lb{LFH}
\eea
Eliminating
the auxiliary fields $H_{\alpha\beta}, \bar H_{\da\db}, C$ by
their algebraic equations of motion
\bea
&& H_{\alpha\beta} =3\,F_{\alpha\beta}~, \; \bar H_{\da\db} = 3\,
\bar F_{\da\db}~,\; C = 0~,
\eea
we arrive at the
standard Maxwell action
\be
L_2(F) = - {1\over2f^2}(F^2 + \bar F^2) = - {1\over 4f^2}{\cal F}^{mn}
{\cal F}_{mn}~, \lb{Maxstand}
\ee
where ${\cal F}_{mn}=\partial_m A_n-\partial_n A_m$.

The basic building-blocks of the $N=3$ BI action are
the analytic superfield strengths. Like in the $N=2$ gauge theory in
$N=2$ HSS \cite{Zu2}, one firstly defines the non-analytic abelian
connections $V^2_1, V^3_2$ via the harmonic zero-curvature equations
\bea
\Dot\Vto -\Dto \Vot =0~,\q \Dth \Vht -\Dht \Vth=0~,\lb{C9}
\eea
where
$V^3_2 = -\widetilde{V^2_1}, \,\delta V^3_2 = iD^3_2 \lambda,\,
\delta V^2_1 = iD^2_1 \lambda$ and the explicit form of the
harmonic derivatives is given in \cite{IZ}. Then
the mutually
conjugated Grassmann-analytic off-shell superfield strengths of the $N=3$
Maxwell theory are constructed as follows \cite{NZ}:
\bea
&& W_{23}={1\over4}(\bar{D}_3)^2V^3_2, \;
\bar{W}^{12}=-{1\over4}(D^1)^2V^2_1.
\lb{2312}
\eea
These off-shell superfield strengths satisfy the following Grassmann
analyticity conditions:
\bea
&& \bDta W_{23}=\bDha W_{23}=\Doa W_{23}=0~, \nn
&& \Doa\bar{W}^{12}=\Dta\bar{W}^{12}=\bDha\bar{W}^{12}=0 \lb{C31b}
\eea
and harmonic differential conditions
\bea
&& \Dth W_{23}=0\q,~\Dot\bar{W}^{12}=0~.\lb{kin}
\eea

We shall need the full off-shell $SU(3)$ singlet component structure of
$W_{23},\bar{W}^{12}$. The explicit expressions for relevant
parts of the latter read
\bea
w_{23} &=& i\theta_2^{\alpha}\theta_3^{\beta} V_\ab
- (\theta_2)^2\theta_3^{\alpha}\bar\theta^{2\db}
\partial_{\db}^{\beta} V_\ab
~, \lb{23single} \\
\bar w^{12} &=& i\bar\theta^{1\da}\bar\theta^{2\db}
\bar V_{\da\db}
+(\bar\theta^2)^2\theta_2^{\alpha}\bar\theta^{1\db}
\partial_{\alpha}^{\da} V_{\da\db}~, \lb{12single}
\eea
where
$$
V_{\alpha\beta} = {1\over 4}\left(
H_{\alpha\beta} + F_{\alpha\beta} \right)~, \;\bar V_{\da\db} =
\ov{(V_{\alpha\beta})}~.
$$
One can directly check that $w_{23}, \bar w^{12}$ on their own
obey the off-shell conditions \p{C31b} and \p{kin}.

The free Maxwell Lagrangian \p{LFH} (with $C=0$), being rewritten
through the newly introduced auxiliary fields $V_{\alpha\beta},
\bar V_{\da\db}$, reads
\bea
&& L_2(F,H,0)\equiv
B_2(F,V)={1\over f^2}[\,V^2+ \bar{V}^2 \nn
&&-\, 2\,(V F+\bar{V}\bar{F})
+{1\over2}(F^2+\bar{F}^2)\,]~. \lb{auxfree}
\eea
The algebraic equations of motion for
$V_{\alpha\beta}, \bar V_{\da\db}$
giving rise to the standard Lagrangian \p{Maxstand} are simply
\bea
&& V_{\alpha\beta} =  F_{\alpha\beta}~, \q \bar V_{\da\db} =
 \bar F_{\da\db}~. \lb{Veqs}
\eea

From the above discussion one infers two important properties of
the off-shell description of $N=3$ gauge theory in $N=3$ HSS
having no direct analogs in the $N=1$ and $N=2$ cases. First, the
free Maxwell component Lagrangian appears in the unusual forms
\p{LFH} or \p{auxfree}, while its standard form is recovered only
after eliminating the auxiliary fields $V_{\alpha\beta}, \bar
V_{\da\db}$ by their linear algebraic equations of motion
\p{Veqs}. Secondly, the off-shell superfield strengths contain just
these tensor auxiliary fields, but not the ordinary gauge field
strengths $F_{\alpha\beta}, \bar F_{\da\db}$.

These surprising features suggest a non-standard approach to
constructing nonlinear and non-polynomial superextensions of the
off-shell $N=3$ Maxwell theory. One should modify
\p{LFH} by proper terms which are nonlinear (and/or
non-polynomial) in the auxiliary fields $V_{\alpha\beta}, \bar
V_{\da\db}$, such that nonlinearities in $F_{\alpha\beta}, \bar
F_{\da\db}$ inherent in the BI action are regained as the result
of eliminating these
auxiliary fields by their {\it nonlinear} equations of motion.
Then one can hope to $N=3$ supersymmetrize the terms nonlinear in
$V_{\alpha\beta}, \bar V_{\da\db}$ with the help of the above
superfield strengths $W_{23}, \bar W^{12}$ which contain just
these auxiliary fields.
\vspace{0.3cm}

\noindent{\it 3.3 $N=3$ BI action}. Without entering into details, the
modification of the free Maxwell Lagrangian \p{auxfree},
such that it becomes the correct bosonic BI Lagrangian,
\bea
&& L_{BI}(F,\bar{F}) ={1\over f^2}\left[1-
\sqrt{-\mbox{det}(\eta_{mn} + {\cal F}_{mn})}\right]
\nn
&&\equiv {1\over
f^2}\left[1-Q(\vp,\bvp)\right]~, Q(\vp,\bvp) = \sqrt{1+X}, \lb{biact}\\
&&
\varphi = F^2~, \;\bar\varphi  = \bar F^2~, \nn && X(\vp , \bvp) \equiv
(\vp+\bvp)+(1/4)(\vp-\bvp)^2~, \lb{not}
\eea
after elimination of the
auxiliary fields $V_{\alpha\beta}, \bar V_{\da\db}$ by their algebraic
equations of motion, is as follows \cite{IZ}
\bea
B(F,V) &=& B_2(F,V) +{1\over f^2}
E(V^2,\bar{V}^2) \nn &=& {1\over f^2}[\,\nu+\bnu - 2(V F +\bar{V}\bar{F})
\nn
&& +\,{1\over2}(\vp+\bvp)+ E(\nu\bnu)]~. \lb{legact}
\eea
Here
$\nu
\equiv V^2, \bnu \equiv \bar V^2$ and $E(\nu\bnu)$
is a real function of the
single argument $\nu\bnu \equiv a$
defined by the following equations
\bea
&& E(a)= 2[2t^2(a)+3t(a)+1]~, \;E(0) = 0, \lb{explE} \\
&& t^4(a) + t^3(a)-{1\over4} a = 0~, \; t(0) = -1~. \lb{quart}
\eea
Note that the generic choice of the function $E(a)$ corresponds to a wide
class of self-dual (by Legendre transformation) extensions of Maxwell
action, the BI one being merely a particular case of these nonlinear actions.
\footnote{See \cite{izup} for a detailed discussion of the self-duality
issues in this new setting.}

The problem of constructing a manifestly $N=3$
supersymmetric superfield action which would yield, in the bosonic sector,
the $F,V$ form \p{legact} of the BI action amounts to setting up a
collection  of superfield monomials which extend the appropriate terms in
the power expansion of the function $E(V^2\bar V^2)$ defined
in \p{explE}, \p{quart}.

Given the function $E(V^2\bar V^2)$, we introduce the new function
$\hat E(V^2\bar V^2)$ by
\bea
&& E(V^2\bar{V}^2)={1\over2}V^2\bar{V}^2\hat{E}(V^2\bar{V}^2)~,
\eea
with $\hat{E}(a)=1-a/4+O(a^2)$. Then the whole sequence of
higher order terms in the $N=3$ generalization of the BI-action
can be written as a closed expression in the analytic superspace,
\be
S_E={1\over32f^2}\int du
d\zeta(^{33}_{11})(W_{23})^2(\bar{W}^{12})^2
\hat{E}(A)~.\lb{nlbiact}
\ee
Here $A$ is the following real
analytic superfield:
 \bea
&&A={1\over2^{11}}(D^1)^2 (\bar{D}_3)^2
[D^{2\alpha} W_{12}D^2_\alpha W_{12} \nn
&&\times\,\bar{D}_{2\da}\bar{W}^{23}
\bar{D}_2^\da\bar{W}^{23}]=V^2\bar{V}^2+\ldots~,\lb{D7}\\
&&W_{12}=D^3_1W_{23}=
-4i\,\theta^{(\alpha}_1\theta^{\beta)}_2V_\ab+\ldots~,\nn
&& \bar{W}^{23}=-D^3_1\bar{W}^{12}~.
\eea

Thus we have obtained an $N=3$ generalization of the Born-Infeld action
using the off-shell Grassmann-analytic potentials $\Vot$ and $\Vth$
\bea
&& S^{N=3}_{BI} = S_2 + S_E~. \lb{BIN3}
\eea
The substitution of generic function $\hat E(A)= 1 +O(A)$ into
\p{nlbiact} yields $N=3$ superextensions of a wide class of
the self-dual nonlinear deformations of Maxwell theory.

Finally, we notice that the above $N=3$ BI action is a minimal
$N=3$ extension of the bosonic BI action. It still remains to examine
whether it admits a treatment as a $N=3$ Goldstone-Maxwell superfield
action describing an off-shell PBGS option $N=6 \rightarrow N=3, d=4$
which could amount on shell to the option $N=8 \rightarrow N=4, d=4$.
By analogy to the situation with $N=2$ BI action \cite{Ke,bik1,bik2},
one could expect that for such an interpretation to be possible
the above action should be modified by some extra terms with
extra $x$-derivatives on them.

\section{CONCLUDING REMARKS}

\noindent In conclusion, it is worth mentioning a few important
unsolved problems in supersymmetric BI theories.

\begin{itemize}
\item
Constructing a BI deformation of the off-shell $N=(1,0), d=6$ super
Maxwell action. Such super BI action should amount to a static gauge
of the space-filling D5-superbrane action in a flat Minkowski background.
\item
Adapting the nonlinear realizations formalism to the case of non-abelian BI
actions. This would seemingly require introducing the notion of generalized
supersymmetry algebras incorporating the non-abelian gauge group structure.
\item
Constructing superconformally invariant versions of $N=2$ and $N=3$ super BI
actions reviewed in this article. Such modifications could play an important
role in the context of the AdS/CFT correspondence \cite{Mald}-\cite{obz},
providing the PBGS form of the effective worldvolume actions of D3-superbranes on
superbackgrounds with the AdS$_n \times S^m$-type bosonic body (AdS$_5\times
S^1$ in the $N=2$ case and AdS$_5\times S^5$ in the $N=3$ and $N=4$ cases).
\end{itemize}

Note that an $N=4$ completion of the $F^4$ term \cite{bkt} of
the $N=2$ superconformal BI action in $N=2$ HSS was found in a recent paper
\cite{biv}. It radically differs from its non-conformal
counterpart \p{N44}:
\bea
&& S_4^{conf} \sim  \int dzdu\,\{\,\ln\frac{{\cal W}}{\Lambda}\,
\ln\frac{\bar{\cal
W}}{\Lambda}\nn
&& +\, (X-1)\frac{\ln (1-X)}{X} + [\, \mbox{Li}_2 (X) -1 \,]
\,\}~. \label{200}
\eea
Here
\bea
&& X= -2\, \frac{q^{+}\cdot q^{-}}{\bar{\cal W}{\cal  W}}~, \label{20}
\eea
and
$$ \mbox{Li}_2 (X) = -\int ^X_0 \frac{\ln (1
-t)}{t}\,dt = \sum_{n=1}^{\infty} \frac{1}{n^2} X^n $$
is Euler dilogarithm ($\Lambda$ is an arbitrary scale). The action \p{200}
is invariant under both $N=2$ superconformal group and $N=4$ supersymmetry
\p{onshell}, hence it is invariant under the whole $N=4$ superconformal group.

As distinct from \p{N44}, the contributions to the component $F^4$ term come
from both the pure ${\cal W}, \bar{\cal W}$ and mixed ${\cal W}, \bar{\cal
W}, q^\pm$ pieces of \p{200}. It reads \bea && \sim \frac{F^2 \bar F^2}{(|
\varphi|^2 + f^{ia}f_{ia})^2}~, \label{boseff} \eea where $\varphi(x),
\bar\varphi(x)$ $(<\varphi> \neq 0)$ and $f^{ia}(x)$ are physical bosonic
fields of the vector $N=2$ multiplet and hypermultiplet. Together they form
a 6-dimensional multiplet of the $R$-symmetry group $SU(4)$ of $N=4$ SUSY.
The $SU(4)$ invariant square of these fields in the denominator of
\p{boseff} ensures the scale and conformal invariance of the corresponding
$x$-space action and can be identified, from the AdS$_5\times S^5$
D3-superbrane perspective \cite{Mald}, with the fifth (radial) co-ordinate
of AdS$_5$.

It would be extremely interesting to find a superconformally invariant
version (if existing) of the $N=3$ BI action \p{BIN3}. Such an action is
expected to be unique and to provide a manifestly off-shell $N=3$
supersymmetric superfield form of the abelian D3-superbrane action on AdS$_5
\times S^5$. It is very important to know this hypothetical maximally
worldvolume supersymmetric form of the D3-brane action both for further
clarifying the AdS/CFT correspondence and, as a closely related goal, for
exploring the precise structure of the quantum low-energy effective action
in $N=4$ SYM theory.

\vspace{0.5cm}
\noindent{\bf Acknowledgements}
\vspace{0.3cm}

\noindent This work was supported by INTAS grant No 00-254,
RFBR grant No 99-02-18417, RFBR-CNRS grant No 01-02-22005 and a grant of
Bogoliubov-Infeld Programme.

\end{document}